\documentclass[journal=jacsat,manuscript=article]{achemso}
\setkeys{acs}{articletitle = true}
\usepackage[version=3]{mhchem}
\usepackage[T1]{fontenc}
\usepackage{graphicx}
\usepackage{epstopdf}
\usepackage{float} 
\usepackage{color}   
\usepackage{multicol} 
\usepackage{array}  
\newcolumntype{L}[1]{>{\raggedright\let\newline\\\arraybackslash\hspace{0pt}}m{#1}}
\mathchardef\mhyphen="2D
\author{Abhishek Pandey}
\affiliation{Ames Laboratory-USDOE and Department of Physics and Astronomy, Iowa State University, Ames, Iowa 50011, United States}
\altaffiliation{Current address: Department of Physics and Astronomy, Texas A\& M University, College Station, Texas 77840, United States}
\email{abhishek.phy@gmail.com}
\author{Saroj L. Samal}
\affiliation{Ames Laboratory-USDOE and Department of Chemistry, Iowa State University, Ames, Iowa 50011, United States}
\altaffiliation{Current address: Department of Chemistry, National Institute of Technology, Rourkela, Odisha 769008, India}
\author{David C. Johnston}
\affiliation{Ames Laboratory-USDOE and Department of Physics and Astronomy, Iowa State University, Ames, Iowa 50011, United States}
\title{CsMn$_4$As$_3$: A layered tetragonal transition-metal pnictide compound with antiferromagnetic ground state}
\begin{document}

\begin{abstract}
We report the synthesis and properties of a new layered tetragonal ternary compound CsMn$_4$As$_3$ (structure: KCu$_4$S$_3$-type, space group: $P4/mmm$, No. 123 and $Z = 2$). The material is a small band-gap semiconductor and exhibits an antiferromagnetic ground state associated with Mn spins. The compound exhibits a signature of a distinct magnetic moment canting event at 150(5)~K with a canting angle of $\approx 0.3^{\circ}$. Although, some features of the magnetic characteristics of this new compound are qualitatively similar to those of the related BaMn$_2$As$_2$, the underlying Mn sublattices of the two materials are quite different. While the Mn square-lattice layers in BaMn$_2$As$_2$ are equally spaced along the $c$-direction with the interlayer distance $d_{\rm L\,Ba} = 6.7341(4)$~\AA, the Mn sublattice forms bilayers in CsMn$_4$As$_3$ with the interlayer distance within a bilayer $d_{\rm L\,Cs} =  3.1661(6)$~\AA\ and the distance between the two adjacent bilayers $d_{\rm B} = 7.290(6)$~\AA. This difference in the Mn sublattice is bound to significantly alter the energy balance between the $J_{1}$, $J_{2}$ and $J_{c}$ exchange interactions within the $J_{1}\mhyphen J_{2} \mhyphen J_{c}$ model compared to that in BaMn$_2$As$_2$ and the other related 122 compounds including the well-known iron-arsenide superconductor parent compound BaFe$_2$As$_2$. Owing to the novelty of its transition metal sublattice, this new addition to the family of tetragonal materials related to the iron-based superconductors brings prospects for doping and pressure studies in the search of new superconducting phases as well as other exciting correlated-electron properties. 
\end{abstract}

\section{Introduction}
Discoveries of high-$T_{\rm c}$ superconductivity in the layered cuprates \cite{Johnston-1997} and 122-, 1111- and 111-type iron-based pnictides \cite{Johnston-2010} suggest that the layered materials containing stacked square lattices of transition metal ions and with inherent antiferromagnetic (AFM) fluctuations/ordering are one of the most promising avenues for the search for new superconductors. After the surprising discovery of high-$T_{\rm c}$ superconductivity in the iron-based compounds the search for new materials that may not necessarily contain iron but crystallize in the related layered structure began and as a result a few new superconductors were discovered \cite{Anand-2013}. BaMn$_2$As$_2$, which is isostructural to the well-known superconductor parent compound BaFe$_2$As$_2$, does not show superconductivity \cite{Singh-2009a, Singh-2009b, An-2009}. Instead, when doped with holes by replacing Ba with K or Rb, BaMn$_2$As$_2$ exhibits half-metallic behavior and a novel magnetic ground state where local-moment AFM of the Mn-sublattice and itinerant ferromagnetism (FM) of the doped holes coexist with each other \cite{Pandey-2012, Lamsal-2013, Pandey-2013, Yeninas-2013, Ueland-2015, Pandey-2015, Ramazanoglu-2016}. Intriguingly, the ordered-moment alignments in these two different types of collinear magnetic orders are orthogonal to each other. Whereas the moments in the G-type AFM of Mn-spins are aligned along the tetragonal $c$-axis, the FM moments of the itinerant holes lie within the $ab$-plane \cite{Pandey-2013, Ueland-2015}. Another consequence of the hole doping in BaMn$_2$As$_2$ is a slow suppression of its N\'eel temperature $T_{\rm N}$ that is reduced from 625~K in the parent compound to 480~K in 40\% hole-doped Ba$_{0.6}$K$_{0.4}$Mn$_2$As$_2$ \cite{Lamsal-2013}. An extrapolation of the $T_{\rm N}$ versus doping concentration $x$ suggests that full suppression of long-range AFM ordering might be achieved at $x \approx 0.8$ \cite{Lamsal-2013}. Thus, by drawing an analogy with the iron-based counterpart BaFe$_2$As$_2$ \cite{Johnston-2010}, one would anticipate the emergence of an interesting ground state in the fully ($x = 1$) hole-doped $A$Mn$_2$As$_2$ compounds ($A$: alkali metal), where these hypothetical phases might be superconducting or at least serve as a potential parent compounds to new superconductors.

\begin{figure*}[h]
\includegraphics[width=6.0in]{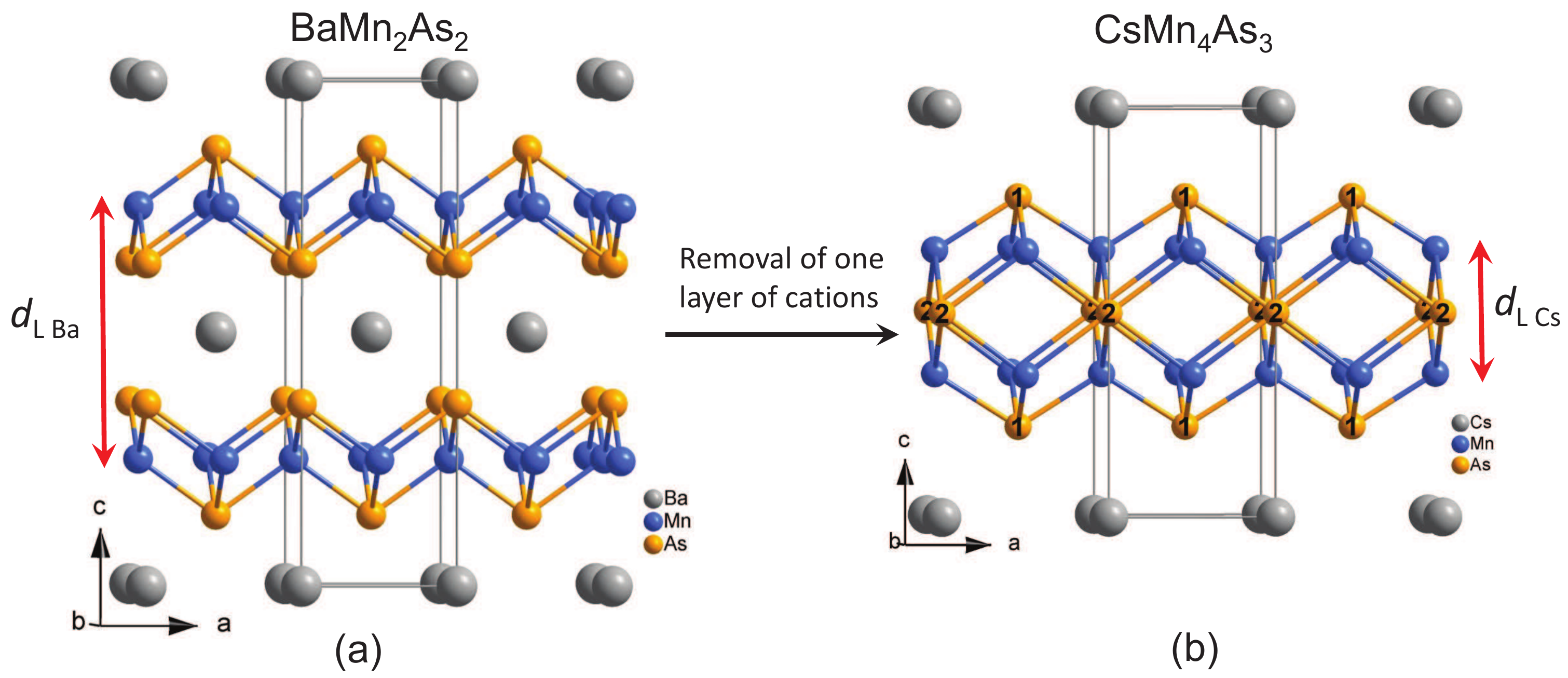}
\caption{Crystal structures of tetragonal compounds (a) BaMn$_2$As$_2$ and (b) CsMn$_4$As$_3$, where the latter is obtained by removing the middle layer of cations from the former. As1 and As2 atoms are indicated in (b) by numbers 1 and 2, respectively. The distance between the two adjacent Mn square-lattice layers are $d_{\rm L\,Ba} = c(1-2z_{\rm Mn}) = 6.7341(4)$~\AA\ in BaMn$_2$As$_2$ and $d_{\rm L\,Cs} = c(1-2z_{\rm Mn}) = 3.1661(6)$~\AA\ in CsMn$_4$As$_3$.}
\label{fig:Structure-1}
\end{figure*}

While our repeated attempts to synthesize the $A$Mn$_2$As$_2$ compounds failed for $A$ = Na, K, Rb and Cs, in our endeavor we discovered a fascinating new tetragonal compound CsMn$_4$As$_3$ which can be best described as a completely collapsed variant of the BaMn$_2$As$_2$ where the entire middle layers of cations and As anions are absent, forming a Mn$_4$As$_3$ bilayer of Mn atoms (Fig.~\ref{fig:Structure-1}). As a result, the ratio of the tetragonal lattice parameters $c/a$ is only 2.44 in CsMn$_4$As$_3$ compared to 3.23 in BaMn$_2$As$_2$. CsMn$_4$As$_3$ is a small band-gap semiconductor that crystallizes in the KCu$_4$S$_3$-type structure\cite{Rudorff-1952, Brown-1980, He-2011} (space group $P4/mmm$, $Z = 2$). This structure is similar to those of the cuprates and the iron-based pnictides, containing a framework of stacked square lattices of transition metal (Mn) ions (Fig.~\ref{fig:Structure-1}). Additionally, our computational and experimental results indicate an AFM ground state in this material. Considering the stimulating observations made in the case of BaMn$_2$As$_2$, this new compound and its plausible variants appear to bring an opportunity to explore some new exciting electronic and/or magnetic ground states possibly including superconductivity.

\section{Experimental and Computational Details}
{\bf Synthesis.} Single crystals of CsMn$_4$As$_3$ were grown by the solution growth technique using MnAs self flux. Cs and MnAs were taken in a ratio of 1:4 in a 2 mL alumina crucible which was then sealed inside a Ta tube under $\sim 1$~atm argon pressure. The Ta tube was then sealed inside a quartz tube under $\sim 1/5$~atm argon pressure. The assembly was then heated to 1230~$^\circ$C in about 20~h, kept there for 5~h and then cooled down to 1110~$^\circ$C in 70~h. At this temperature the excess flux was decanted using a centrifuge. Several platelike shiny crystals of typical size $5\times5\times0.5$~mm$^3$ were obtained from the growth (Inset, Fig.~1S, supporting information).
 
\begin{figure*}[h]
	\includegraphics[width=6.0in]{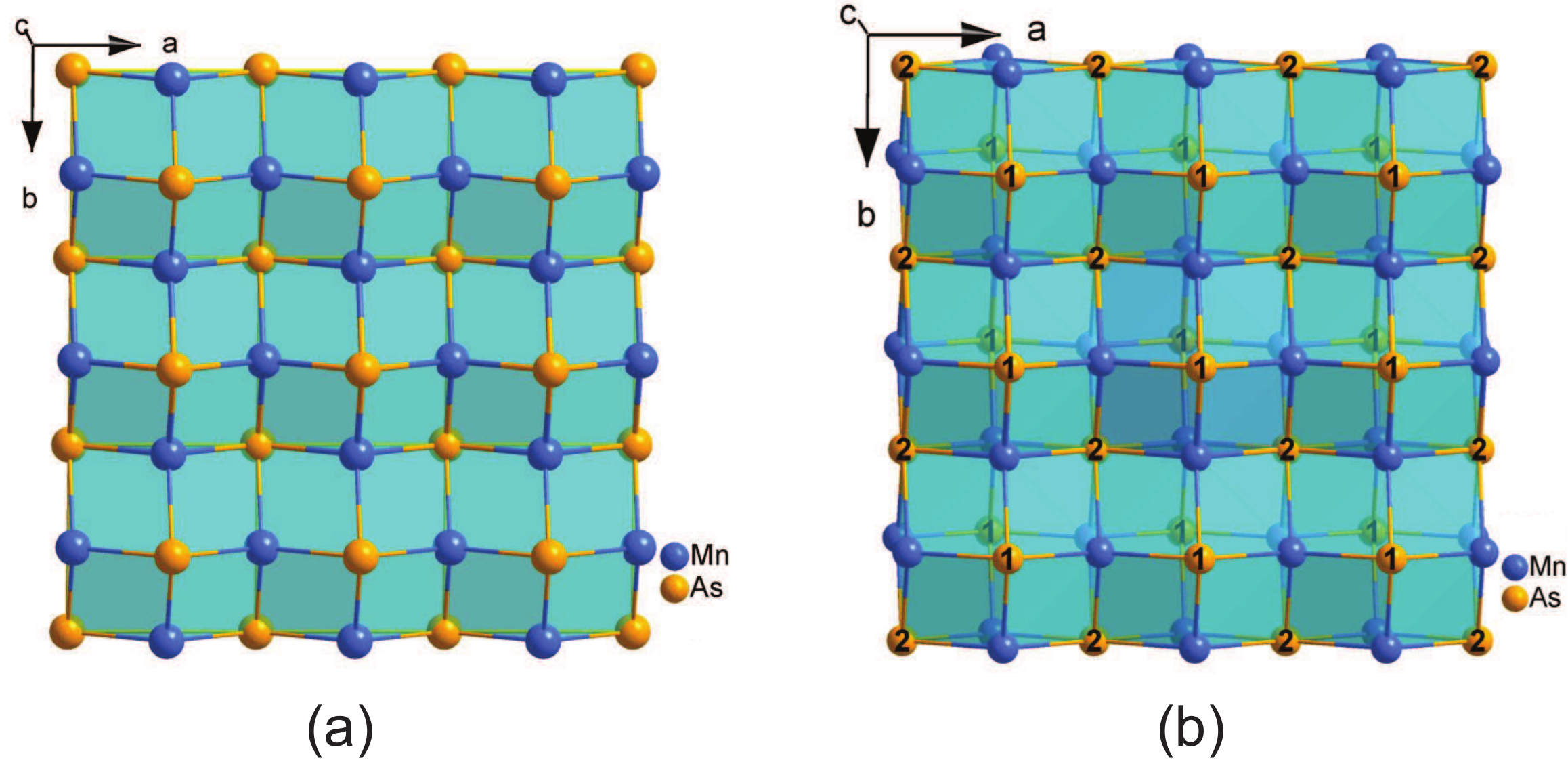}
	\caption{Arrangement of (a) [Mn$_2$As$_2$]$^{-2}$ layer in BaMn$_2$As$_2$ and (b) [Mn$_4$As$_3$]$^{-1}$ bilayer in CsMn$_4$As$_3$. As1 and As2 atoms are indicated in (b) by numbers 1 and 2, respectively.}
	\label{fig:Structure-2}
\end{figure*}

{\bf X-ray Diffraction Studies.} Powder x-ray diffraction (XRD) data were collected at room temperature using a Rigaku Geigerflex diffractometer utilizing Cu-$K_\alpha$ radiation ($\lambda = 1.54059$~\AA). Rietveld refinement was performed using the FullPROF package \cite{Carvajal-1993} (Fig.~S1, supporting information).

Single crystals of CsMn$_4$As$_3$ were sealed in glass capillaries inside a N$_2$-filled glovebox. Single-crystal diffraction data sets were collected at room temperature over a $2\theta$ range of $\sim 6^{\circ}$ to $\sim 60^{\circ}$ with $0.5^{\circ}$ scans in $\omega$ and 10~s per frame exposures with the aid of a Bruker SMART CCD diffractometer equipped with Mo-$K_\alpha$ radiation ($\lambda = 0.71073$~\AA). The data showed a primitive-tetragonal lattice and the intensity statistics indicated a centrosymmetric space group. The reflection intensities were integrated with the APEX II program in the SMART software package. \cite{SMART-1996} Empirical absorption corrections were employed using the SADABS program. \cite{Blessing-1995} The space group $P4/mmm$ (No. 123) of the structure was determined with the help of XPREP and SHELXTL 6.1.\cite{SHELXTL-2000} The structure was solved by direct methods and subsequently refined on $|F^{2}|$ with a combination of least-square refinements and difference Fourier maps. 

The refinements of CsMn$_4$As$_3$ converged to $R_1 = 0.0709$, $R_{\rm W} = 0.0973$ for all data with goodness of fit 1.102 and maximum residuals of 3.92 and $-1.68$~e/\AA$^3$ that were 0.8 and 1.1~\AA\ from the Cs and As2 sites, respectively. The refinement parameters and some crystallographic data are given in Table~\ref{tab:Table-1}. The corresponding atomic positions and significant bond distances are listed in Table~\ref{tab:Table-2} and Table~\ref{tab:Table-4}, respectively. The anisotropic displacement parameters are tabulated in Tables~S1. The cif files are provided in the Supporting Information. 

{\bf Magnetic, Thermal and Electrical Transport Measurements.} Temperature $T$- and magnetic field $H$-dependent magnetic measurements were performed on an oriented single crystal of CsMn$_4$As$_3$ using a Quantum Design, Inc. (QDI), superconducting quantum interference device magnetometer. Heat capacity $C_{\rm p}$ versus $T$ measurements were performed on a single crystal sample employing QDI's physical properties measurement system (PPMS) in two parts--first between 1.8 to 300~K using the N-grease and then between 290 to 344~K on the same crystal using the H-grease, where both N- and H-grease were supplied by QDI. In-plane electrical resistivity $\rho_{\rm ab}$ versus $T$ measurements were performed on a single crystal sample using the PPMS over the $T$ range of 1.8-300~K\@.

{\bf Electronic Structure Calculations.}
Electronic structure calculations on CsMn$_4$As$_3$ were performed self-consistently using the tight-binding linear-muffin-tin-orbital (TB-LMTO) method within the atomic sphere approximation \cite{Krier-1995}. The simplest model that allows the AFM coupling was employed, which is in space group $P4mm$ with two independent Mn atoms. The exchange and correlation were treated in the local spin-density approximation (LSDA) calculations. Scalar relativistic corrections were included.\cite{Koelling-1977} The Wigner-Seitz radii of the spheres were assigned automatically so that the overlapping potentials would be the best possible approximation to the full potential.\cite{Jepsen-1995} The radii (\AA) were: Cs = 2.39, Mn1 = Mn2 = 1.38, As1 = 1.61, As2 = As3 = 1.48. No additional empty spheres were needed subject to an 18\% overlap restriction between atom-centered spheres. Basis sets of Cs $6s, (6p, 5d, 4f)$; Mn $4s, 4p, 3d$; and As $4s, 4p, (4d)$ (downfolded orbitals in parentheses) were employed, and the tetrahedron method using a $18\times18\times6$ mesh of $k$-points was applied to perform the reciprocal space integrations.

For bonding analysis, the crystal orbital Hamilton populations (COHPs)\cite{Dronskowski-1993} of all filled electronic states for selected atom pairs were calculated. The weighted integration of COHP curves of these atom pairs up to the Fermi energy ($E_{\rm F}$) provides ICOHP values, i.e., total (integrated) Hamilton populations  (Table~\ref{tab:Table-4}), which are approximations of relative bond strengths. The COHP analyses provide the contributions of the covalent parts of particular pair-wise interactions to the total bonding energy of the crystal. 

\section{Results and Discussion}
{\bf Structure.} 
A new layered tetragonal compound CsMn$_4$As$_3$ has been discovered which crystallizes in the primitive-tetragonal space group $P4/mmm$ (No. 123 and $Z = 2$) and has a framework of [Mn$_4$As$_3$]$^{-}$ layers and Cs-spacer atoms between the layers. CsMn$_4$As$_3$ can be considered as a derivative of the well-known G-type AFM compound BaMn$_2$As$_2$ ($\equiv$ Ba$_2$Mn$_4$As$_4$) that contains two layers of [Mn$_2$As$_2$]$^{-2}$ with a cation spacer layer in between. When replacing Ba with Cs, one layer of cations is entirely removed and the two anion layers collapse to form a single [Mn$_4$As$_3$]$^{-}$ layer in the resultant 143-compound CsMn$_4$As$_3$ (Fig.~\ref{fig:Structure-1}). As a result, the $c$-axis is compressed to about 3/4 of the $c$-axis of BaMn$_2$As$_2$. This altered structure seems to primarily result from two synergetic effects--(i) charge balance between the cations and the anion, where the formal charge on CsMn$_4$As$_3$ can be assigned according to the Zintl-Klemm rule by assuming complete electron transfer from cations to anions, yielding (Cs$^{+}$)[(Mn$^{+2}$)$_4$(As$^{-3}$)$_3$]; and (ii) a weaker cation-anion interaction in CsMn$_4$As$_3$ compared to that in the related compound BaMn$_2$As$_2$ \cite{Samal-2013a} as discussed in the "Electronic Structure and Chemical Bonding section" below. 

The basic building block in the CsMn$_4$As$_3$ structure is the Mn$_8$ rectangular polyhedron in which all the faces are capped by As atoms leading into a Mn$_8$@As$_6$ unit. This arrangement results as a consequence of the collapse of the two [Mn$_2$As$_2$]$^{-2}$ layers into a single [Mn$_4$As$_3$]$^{-}$ layer. The face-capped Mn$_8$@As$_6$ polyhedra are connected through Mn--Mn and Mn--As bonds. In contrast, the basic building units in the BaMn$_2$As$_2$ structure are the square networks of Mn atoms capped by As atoms. These Mn$_4$ blocks are edge-shared in the $ab$-plane to form a 2D layer with each Mn$_4$ square capped by an As atom in a zigzag manner (Fig.~\ref{fig:Structure-2}). 

\begin{table*}
  \caption{Crystal data and the structural refinement parameters for CsMn$_4$As$_3$ that crystallizes in a tetragonal structure with $P4/mmm$ (No. 123) space group symmetry and $Z = 2$ formula units/cell.}
  \label{tab:Table-1}
  \begin{tabular}{ll}
	 \hline
	  Parameter & Estimated value/range \\
	 \hline
    Formula weight (g) &	577.42 \\
		Unit cell dimensions (\AA)	& $a = 4.2882(2)$ and $c = 10.456(2)$ \\
		Volume (\AA$^3$) &	$192.3(1)$ \\
		Density (g/cm$^3$) & 4.98 \\
		Absorption coefficient (mm$^{-1}$) & 132.63\\ 
		Theta range (degree)	& 1.98 to 28.998 \\
		Index ranges &	$-4 \le h \le 5$; $-4 \le k \le 5$; $−13 \le l \le 13$ \\
		Reflections collected &	1725 \\
		Independent reflections &	194 \\ 
		Data/parameters &	194/12 \\ 
		Goodness-of-fit on $F^2$ &	1.102 \\
		Final $R$ indices [$I > 2\sigma(I)$]\textsuperscript{\emph{a}} &	$R_1 = 0.0593$; $wR_2 = 0.0865$ \\ 
		$R$ indices (all data) &	$R_1 = 0.0709$; $wR_2 = 0.0973$ \\ 
		Largest diff. peak and hole (e \AA$^{-3}$) &	3.92 [0.81~\AA~from Cs] and $-1.68$ [1.16~\AA~from As2] \\
		\hline
   \end{tabular}
	
  \textsuperscript{\emph{a}}$R = \Sigma||F_{\rm o}| - |F_{\rm c}||/\Sigma|F_{\rm o}|; R_{w} = [\Sigma w(|F_{\rm o}|-|F_{\rm c}|)^{2}/\Sigma w(F_{\rm o})^{2}]^{1/2}; w = 1/\sigma F^{2}.$
\end{table*}

There are two types of arsenic atoms present in CsMn$_4$As$_3$---the bridging As atoms (As1) that are bonded to eight Mn atoms in a rectangular geometry and the terminal As atoms (As2) that are bonded to four Mn atoms in a similar way as in BaMn$_2$As$_2$ which has only one type of As atom (Fig.~\ref{fig:Structure-1}). In CsMn$_4$As$_3$, each Mn is bonded to four As in a distorted tetrahedron geometry (Fig.~S2, supporting information) with a longer bond length (2.666~\AA) between Mn and shared As (As1) along the $c$-direction and a shorter bond length (2.533~\AA) between Mn and the terminal As (As2). The As-Mn-As bond angle ranges from $107.08(1)^{\circ}$ to $115.65(1)^{\circ}$, significantly deviating from an ideal tetrahedral geometry. On the other hand, in the case of BaMn$_2$As$_2$, each Mn is bonded to four As in a slightly distorted tetrahedron (As-Mn-As bond angle $108.64^{\circ}-109.89^{\circ}$) with equal Mn--As bond distances (2.566~\AA) \cite{Johnston-2010}. 

\begin{table*}
  \caption{Wyckoff positions, site point symmetry, atomic coordinates and isotropic equivalent displacement parameters for CsMn$_4$As$_3$.}
  \label{tab:Table-2}
  \begin{tabular}{lllllll}
    \hline
   	atom &	Wyckoff &	site & $x$ &	$y$ &	$z$ &	$U_{\rm eq}$(\AA$^2$)\textsuperscript{\emph{a}} \\
    \hline
		Cs & $1b$ &	$4/mmm$	& 0 & 0 & 0	& 0.023(1) \\
		Mn & $4i$ &	$2mm$ &	0 &	1/2 &	0.3486(2)	& 0.019(1) \\
		As1 &	$1a$ & $4/mmm$ & 0 &	0 & 1/2 &  0.022(1) \\
    As2 &	$2h$ & $4mm$ & 1/2	& 1/2	& 0.2202(2) &	0.012(1) \\
   \hline
  \end{tabular}
	
	\textsuperscript{\emph{a}}$U_{eq}$ is defined as one-third of the trace of the orthogonalized $U^{ij}$ tensor.
\end{table*}

\begin{table*}
  \caption{Structural features and parameters, Pauling's metallic radii $r_{A(12)}$, $A$--As distance $d_{A-\rm{As}}$ ($A$ = Ba, Cs and Ca), Mn/Fe--As distances $d_{\rm Mn/Fe-As}$ and N\'eel temperature $T_{\rm N}$ for BaMn$_2$As$_2$, CsMn$_4$As$_3$ and CaFe$_4$As$_3$.}
  \label{tab:Table-3}
  \begin{tabular}{L{3.0cm} | L{3.8cm} | L{3.8cm} | L{4.0cm}}
    \hline
  	 & BaMn$_2$As$_2$	& CsMn$_4$As$_3$	 &	CaFe$_4$As$_3$\\
    \hline
		Structure \& Space group & Tetragonal, $I4/mmm$ & Tetragonal, $P4/mmm$ & Orthorhombic, $Pnma$ \\
		\hline
		Structural Features & Layered structure that contains [Mn$_2$As$_2$]$^{-2}$ layers with cation layer as spacer. & Layered structure that contains [Mn$_4$As$_3$]$^{-1}$ layers with cation layer as spacer. & 3D channel/tunnel like structure where cations fill the channel made of Fe and As. \\
	 & Ba has cubic coordination by As. &  Cs has cubic coordination by As. & Ca has trigonal prismatic coordination \\
		\hline
		Unit cell dimensions (\AA) & $a = 4.1684(2)$ $c = 13.4681(8)$ $c/a = 3.2309(4)$  & $a = 4.2882(2)$ $c = 10.456(2)$ $c/a = 2.4383(6)$ & $a = 11.918(2)$ $b = 3.749(1)$ $c = 11.625(3)$ $a/b = 3.179(1)$ $c/b = 3.101(2)$\\
		\hline
		$z_{\rm Mn}$ & 1/2 & 0.3486(2) & -- \\
		\hline
		$r_{A(12)}$~(\AA) & 2.215 & 2.67 & 1.97 \\
		\hline
		$d_{A-{\rm As}}$~(\AA) & 3.492 & 3.803 & 2.969-3.152 \\
		\hline
		$d_{\rm Mn/Fe-As}$~(\AA) & 2.566 & 2.533 and 2.666 & 2.380-2.465 (CN = 4), \hfill 2.427-2.610 (CN = 5) \\
\hline
$T_{\rm N}$(K) & 625 & $>300$ & 90 ($T_{\rm N1}$) and 26 ($T_{\rm N2}$)  \\
\hline
  \end{tabular}
\end{table*}

In the related iron-based analogue 143-compound CaFe$_4$As$_3$ that crystallizes in an orthorhombic structure (space group: $Pnma$), Fe atoms are in two different coordination environments---a strongly distorted tetrahedron and a square pyramid, corresponding to two different sets of Fe--As bond lengths \cite{Todorov-2009}. The compound has a 3D open framework containing a covalent channel consisting of Fe and As. The Ca$^{+2}$ cations fill the channel that runs along the $b$-axis of the orthorhombic cell.  Thus, although CsMn$_4$As$_3$ and CaFe$_4$As$_3$ have the same stoichiometry, they are structurally entirely different and their properties cannot be compared on the same footing (Table~\ref{tab:Table-3}). 

\begin{figure}
	\includegraphics[width=3.0in]{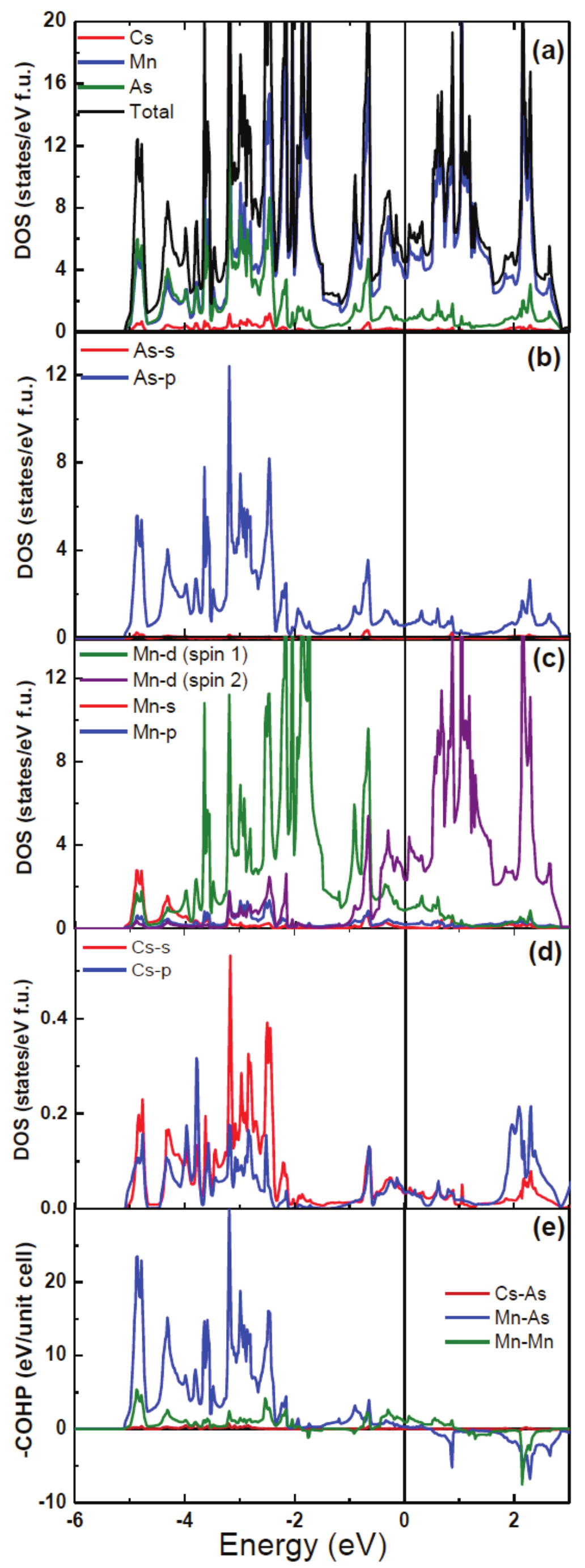}
	\caption{Results of spin-polarized LMTO-ASA calculations for CsMn$_4$As$_3$. (a) Total and atom decomposed densities of states (DOS). Partial projections of orbital DOS are shown for (b) As $4s$ and $4p$; (c) Mn $4s$, $4p$ and spin-polarized $3d$ (d) Cs $6s$ and 6p. (e) COHP values for Cs--As, Mn--As and Mn--Mn. The DOS's plotted in figures (a) to (d) are for both spin directions. In all five panels the Fermi energy is set to zero.}
	\label{fig:DOS}
\end{figure}

{\bf Electronic Structure and Chemical Bonding.}
Figures~\ref{fig:DOS}(a-d) show the total electronic density of states (DOS) along with the individual atom and orbital projections of the DOS for CsMn$_4$As$_3$ obtained from the tight-binding LMTO calculations using the LSDA (both majority and minority spin states are combined within the diagrams). The calculations were carried out for an A-type collinear AFM model where the Mn spins are aligned along the $c$-axis within the space group $P4mm$ (Fig.~S3, supporting information). The COHP curves for each pair-wise interaction as a function of energy are shown in Fig.~\ref{fig:DOS}(e). The $E_{\rm F}$ lies in a nonzero DOS region of ${\cal D}^{\rm band}(E_{\rm F}) \sim 5$~states/eV f.u. for both spin directions. This suggests a metallic character of the compound. The As $4s$ states lie below $-11$~eV and are not shown in the DOS plot. The As $4p$ states appear mostly in the $-5.2$ to 2.0~eV range. The Mn $3d$ states are spin polarized with a zero-temperature ordered magnetic moment of 3.50~$\mu_{\rm B}$ per Mn atom. This value of the magnetic moment of Mn occurs because the majority spin states are nearly filled (4.5 $e^{-}$/Mn) and the minority spin states are partially filled (1.0 $e^{-}$/Mn) [Fig.~\ref{fig:DOS}(c)]. 

\begin{table*}
  \caption{Bond lengths, average molar integrated crystal orbital Hamilton population (ICOHP) values, number of bonds per cell $n$ and percentage contribution of each type of bond to the total bonding per unit cell for CsMn$_4$As$_3$ obtained from the spin polarized calculations.}
  \label{tab:Table-4}
  \begin{tabular}{lccccc}
    \hline
  	Bond & Length(\AA) & $-$ICOHP/bond(eV)	& $n$ &	$-$ICOHP/cell	& \% \\
    \hline
    Mn1--As1	& 2.533(1) & 2.44 &	4	& 9.76 &  25.9\\
		Mn2--As2	& 2.533(1) & 2.44 &	4	& 9.76 & 25.9 \\
		Mn1--As3	& 2.665(1) & 1.59	& 4	& 6.36 & 16.9 \\
		Mn2--As3	& 2.665(1) & 1.59	& 4	& 6.36 & 16.9 \\
		Mn1--Mn1	& 3.032(2) & 0.55	& 4	& 2.2 & 5.8 \\
		Mn2--Mn2	& 3.032(2) & 0.55	& 4	& 2.2 & 5.8 \\
		Mn1--Mn2	& 3.170(2) & 0.27	& 2	& 0.54 & 1.4 \\
		Cs--As	& 3.802(2) & 0.07 &	8	& 0.56 & 1.5 \\
		\hline
  \end{tabular}
\end{table*}

Chemical bonding features can be judged to some extent from the DOS and COHP curves as well as the integrated crystal orbital Hamilton population (ICOHP) data. The important bond distances and the average $-{\rm ICOHP}$ values for each bond type and their percent contributions to the total polar-covalent bonding per unit cell are listed in Table~\ref{tab:Table-4}. The individual heteroatomic Mn--As orbital interactions have $-{\rm ICOHP}$ values of 2.44~eV and 1.59~eV for Mn--As2 and Mn--As1, respectively, which are significantly larger than all other contacts (Table~\ref{tab:Table-4}). The bonding contributions of Mn--As and Mn--Mn contacts are ~85\% and ~13\%, respectively, making them the most dominant contributors to the total bonding energy. The $-{\rm ICOHP}$ value of individual polar heteroatomic Cs--As orbital interactions is 0.07~eV, which contributes only about 2\% to the total bonding energy. This indicates a very weak Cs--As interaction in CsMn$_4$As$_3$, which is likely one of the reasons for the absence of the intermediate cation layer in this compound. A similar weak cation-anion interaction was observed in Na$_6$Au$_7$Cd$_{16}$,\cite{Samal-2011} which is closly related to Ca$_6$Pt$_8$Cd$_{16}$\cite{Samal-2013a}. In the latter compound a strong cation-anion (Ca--Pt) bonding was observed due to the availability of $3d$ orbitals of Ca which are naturally lacking in Na. To compare the bond energies of the $A$-As ($A$ = Ba or Cs) bonds in the presently reported and the BaMn$_2$As$_2$ structure types, we have performed calculations on Cs$T_4$As$_3$ ($T$ = Mn, Zn and Cd) and BaMn$_2$As$_2$ compounds (Table~S2, supporting information). The calculations show that the $-$ICOHP value for the Ba-As bond is double than that of the Cs-As bonds, which indicates a weaker cation-anion interaction in Cs$T_4$As$_3$ compared to that in the BaMn$_2$As$_2$. In addition, the number of such bonds per unit cell in BaMn$_2$As$_2$ are double the number of those bonds in CsMn$_4$As$_3$. Hence the bond energy contribution of the $A$-As interaction per unit cell toward the structure stability in the later structure is four times smaller in magnitude than in the former. 

Mn--As bonding in CsMn$_4$As$_3$ arises largely from interactions between Mn 4$s$, 3$d$ and As 4$p$ orbitals, whereas Mn--Mn bonding comes from Mn 4$s$, 4$p$ and Mn 4$s$, 4$p$ orbital interactions. It may be noted that Mn 3$d$ along with the 4$s$ and 4$p$ orbitals significantly participate in the chemical bonding with As and thus result in a more delocalized 3$d$ state of Mn compared to our earlier-reported Mn-based intermetallic compounds.\cite{Samal-2013c, Samal-2014}

\begin{figure}[h]
	\includegraphics[width=3.3in]{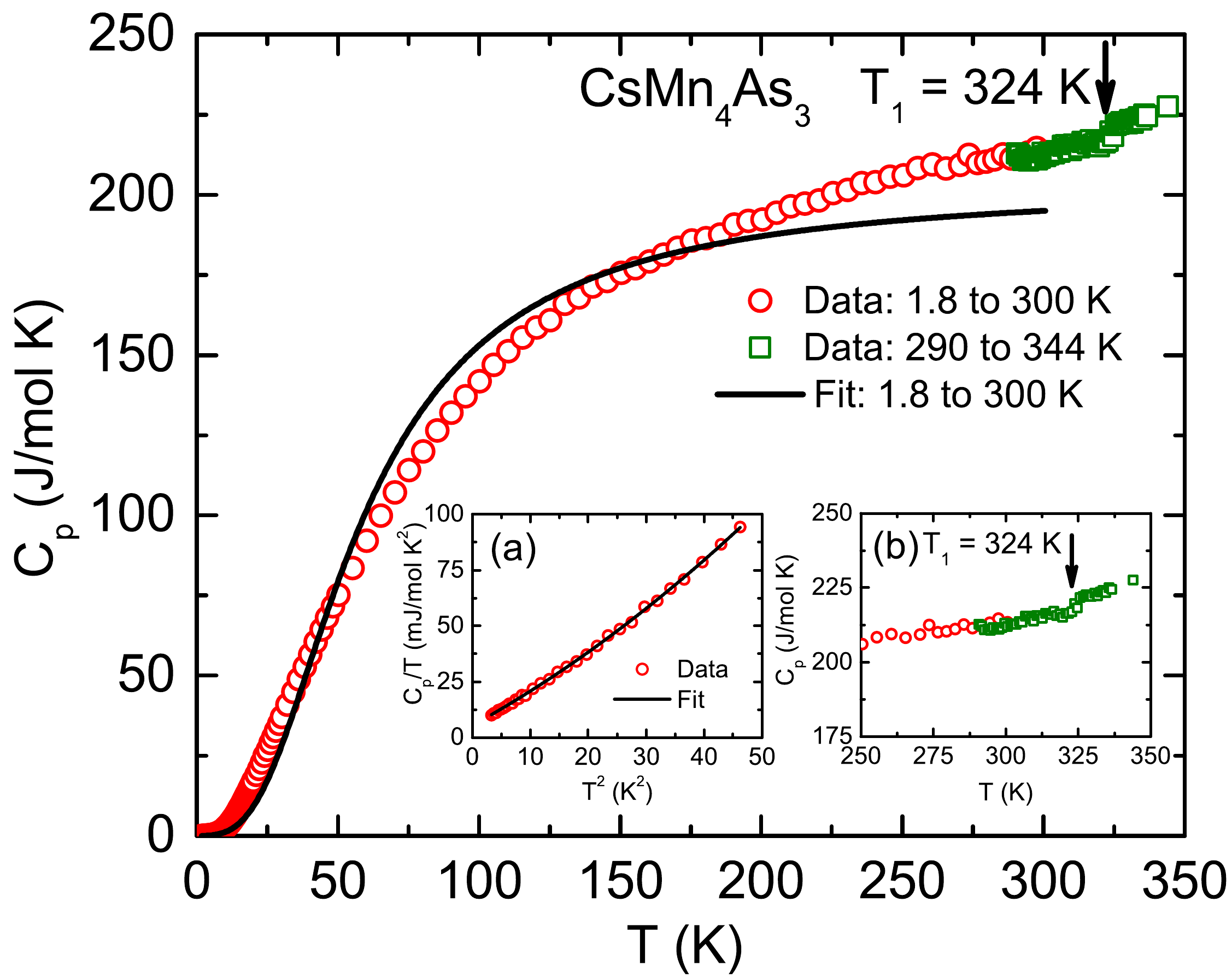}
	\caption{Heat capacity $C_{\rm p}$ versus temperature $T$ for CsMn$_4$As$_3$. Open red circle and open green squares are the data taken between 1.8 to 300~K and between 290 to 344~K, respectively. Vertical arrow indicates the transition temperature $T_1$. Solid curve is a fit using the Debye model to the data between 1.8 to 300~K\@. Inset (a): Magnified view of the $C_{\rm p}(T)$ data near the transition temperature $T_1$. Inset (b): variation of $C/T$ with $T$ at low temperature. Solid curve is a fit using the eq~\ref{eq:HC-LT} described in the text.}
	\label{fig:HC}
\end{figure}

{\bf Heat Capacity.}
Figure~\ref{fig:HC} shows the temperature $T$ variation of the heat capacity $C_{\rm p}$ of CsMn$_4$As$_3$. The $C_{\rm p}$ at $T = 300$~K is 213.4~J/mol~K\@, which is about 7\% larger than the expected Dulong-Petit value $C_{\rm V} = 3nR = 199.6$~J/mol~K, where $R$ is the molar gas constant and $n$ is the number of atoms per formula unit which is 8 for CsMn$_4$As$_3$. To investigate if this discrepancy originates from a large electronic contribution to the $C_{\rm p}$, we first focus at the low-$T$ data. At low temperatures, in absence of any magnetic contribution or phase transition the $C_{\rm p}(T)$ follows
\begin{equation}
C_{\rm p} = \gamma T + \beta T^3 + \delta T^5,
\label{eq:HC-LT}
\end{equation}
where $\gamma$ is the electronic Sommerfeld coefficient, and $\gamma T$ and ($\beta T^3 + \delta T^5$) represent the low-$T$ electronic and lattice contributions to the $C_{\rm p}$, respectively. We get a good fit to the data using eq~\ref{eq:HC-LT} [Inset (a), Fig.~\ref{fig:HC}] and the fitted values of the coefficients are $\gamma = 5.4(3)$~mJ/mol~K$^2$, $\beta = 1.43(3)$~mJ/mol~K$^4$ and $\delta = 1.05(7)\times 10^{-2}$~mJ/mol~K$^6$. Using the estimated $\gamma$, the electronic contribution to the $C_{\rm p}$ at 300~K is calculated as 1.62~J/mol~K\@. This is much smaller than the observed enhancement over the Dulong-Petit value, and thus rules out an electronic origin of it. Additionally, our attempt to fit the $C_{\rm p}(T)$ of CsMn$_4$As$_3$ between 1.8 to 300~K using the Debye model\cite{Gopal-1966,Goetsch-2012} of acoustic lattice phonons resulted in a poor fit to the data (Fig.~\ref{fig:HC}). These two observations together suggest the presence of a sizable magnetic contribution to $C_{\rm p}(T)$ of CsMn$_4$As$_3$ near room temperature, indicating that $T_{\rm N} > 300$~K\@. A similar observation was made earlier in the case of Fe-doped BaMn$_2$As$_2$ \cite{Pandey-2011}. High-$T$ $C_{\rm p}(T)$ data of CsMn$_4$As$_3$ taken between 290 to 344~K exhibit a jump of about 3~J/mol~K at $T_{1} = 324$~K [Inset (b), Fig.~\ref{fig:HC}], which is identified below as a likely magnetic transition temperature where one of the two magnetic moment canting events takes place.

The nonzero value of $\gamma$ suggests the presence of conduction electron states at $E_{\rm F}$ of CsMn$_4$As$_3$. The DOS at the $E_{\rm F}$ ${\cal D}^\gamma(E_{\rm F})$ can be estimated from the experimentally estimated $\gamma$ using 
\begin{equation}
{\cal D}^\gamma(E_{\rm F}) \left[{\frac{\rm states}{\rm eV~f.u.}} \right] = \frac{1}{2.359}\gamma\left[\frac{\rm mJ}{\rm mol~K^{2}}\right],
\label{eq:DOS}
\end{equation} 
where ${\cal D}^\gamma(E_{\rm F})$ is for both spin directions. The calculated value of ${\cal D}^\gamma(E_{\rm F})$ is 2.3(1)~states/eV~f.u. for both spin directions. A similar ${\cal D}^\gamma(E_{\rm F})$ was observed in the lightly (<5\%) hole-doped BaMn$_2$As$_2$\cite{Pandey-2012}. The ${\cal D}^\gamma(E_{\rm F})$ value is about a factor of two smaller than obtained from the above band structure calculations, suggesting the presence of localization effects in this layered material. From the experimentally determined $\beta$, the Debye temperature $\Theta_{\rm D}$ of CsMn$_4$As$_3$ is calculated as $\Theta_{\rm D} = (12\pi^4Rn/5\beta)^{1/3} = 222(2)$~K\@. The estimated magnetic entropy $S_{\rm mag}$ that approaches $R{\rm ln}6$ at high temperatures [Fig.~S4(b), supporting information] suggests the presence of high-spin $S = 5/2$ state of the Mn$^{+2}$ ions in this compound.  

\begin{figure}[t]
	\includegraphics[width=3.3in]{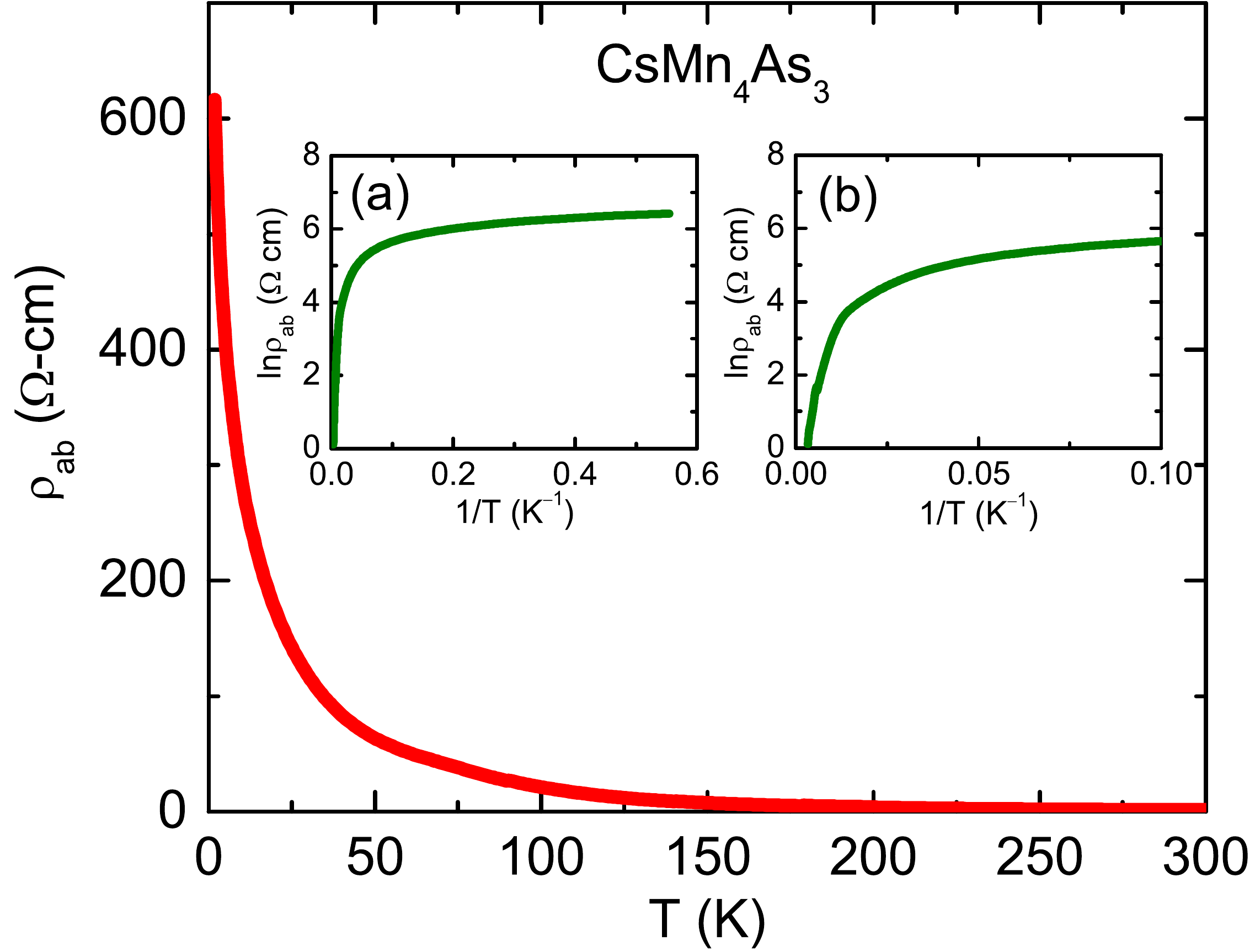}
	\caption{Electrical resistivity $\rho_{ab}$ measured in the basal $ab$-plane versus temperature $T$ of CsMn$_4$As$_3$. Inset(a): resistivity on a logarithmic scale ln$\rho_{ab}$ versus $1/T$. Inset (b): ln$\rho_{ab}$ versus $1/T$ in a restricted $1/T$ range.}
	\label{fig:Res}
\end{figure}

{\bf Electrical Resistivity.}
The electrical resistivity within the basal plane $\rho_{ab}$ versus $T$ data of CsMn$_4$As$_3$ are shown in the Fig.~\ref{fig:Res}. In agreement with the $C_{\rm p}(T)$ data and the band calculations discussed above the $\rho_{ab}$ has a finite value at the lowest $T$ of our measurement, suggesting a non-insulating ground state in this compound. However, the very high value of $\rho_{ab~2~{\rm K}} = 616.2~\Omega~{\rm cm}$, which is several orders of magnitude larger than the resistivities of normal metals, indicates that the material is not a good metal either. While the $\rho_{ab}$ shows a strong $T$-dependence and monotonically decreases to a value of $1.4~\Omega~{\rm cm}$ at 300~K\@, it does not show the activated behavior expected in a band semiconductor as evident from the absence of any linear region in ln$\rho_{\rm ab}$ versus $1/T$ plots [Insets (a) and (b), Fig.~\ref{fig:Res}]. These observations collectively suggest that the mechanism of electrical transport in CsMn$_4$As$_3$ is qualitatively similar to that in the related small band-gap semiconductor BaMn$_2$As$_2$, which also shows a non-metallic $T$-dependence and a finite sample-dependent $\rho_{ab}$ at $T \rightarrow 0$ \cite{Singh-2009a, An-2009, Pandey-2012}. The observed $T$-dependence of $\rho_{ab}$ likely originates from the limited availability of current-carrying $s$ and $p$ states at the $E_{\rm F}$ of CsMn$_4$As$_3$. As evident from the band calculations (Fig.~\ref{fig:DOS}) the contribution of the $s$ and $p$ states to the ${\cal D}(E_{\rm F})$ is only $\sim 1$~state/eV f.u. for both spin directions, whereas the heavier $d$ states that may not take part in the conduction contribute $\sim 4$~states/eV f.u.\ for both spin directions. From the observations made in the heat capacity and the electrical resistivity measurements as well as from the results of the band calculations it appears that bad-metal type electrical conduction behavior of CsMn$_4$As$_3$ is an outcome of limited availability of conduction carriers and strong electronic correlations. It appears that similar to the hole-doped BaMn$_2$As$_2$, this material can likely be pushed into a metallic ground state by hole/electron doping and/or by the application of external pressure \cite{Satya-2011}. We have also evaluated the possibility of the variable range hopping \cite{Mott-1969, Shklovskii-1984}-type electrical conduction in our material. However, the data do not seem to follow this behavior, especially at the low temperatures (Fig.~S5, supporting information).   

\begin{figure}[t]
	\includegraphics[width=3.3in]{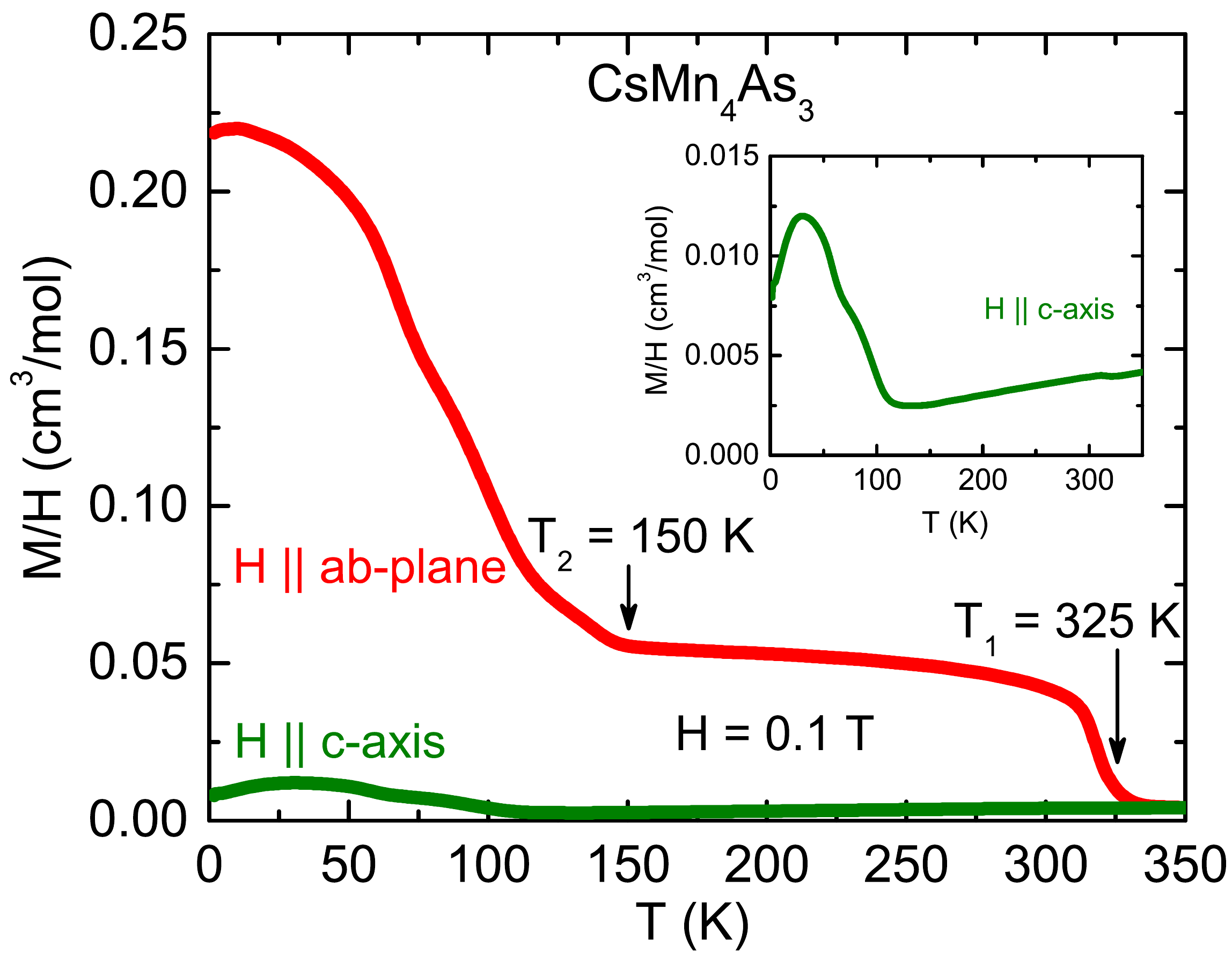}
	\caption{Anisotropic magnetic susceptibilities $\chi_{ab,c} \equiv M_{ab,c}/H$ versus temperature $T$ of CsMn$_4$As$_3$ measured at applied dc magnetic fields of $H = 0.1$~T\@. Vertical arrows indicate the two transitions observed in $\chi_{ab}$ at $T_{1} = 325(5)$~K and $T_{2} = 150(5)$~K, respectively. Inset:  expanded plot of the $\chi_{c}(T)$.}
	\label{fig:MT}
\end{figure} 
\begin{figure*}
	\includegraphics[width=3.3in]{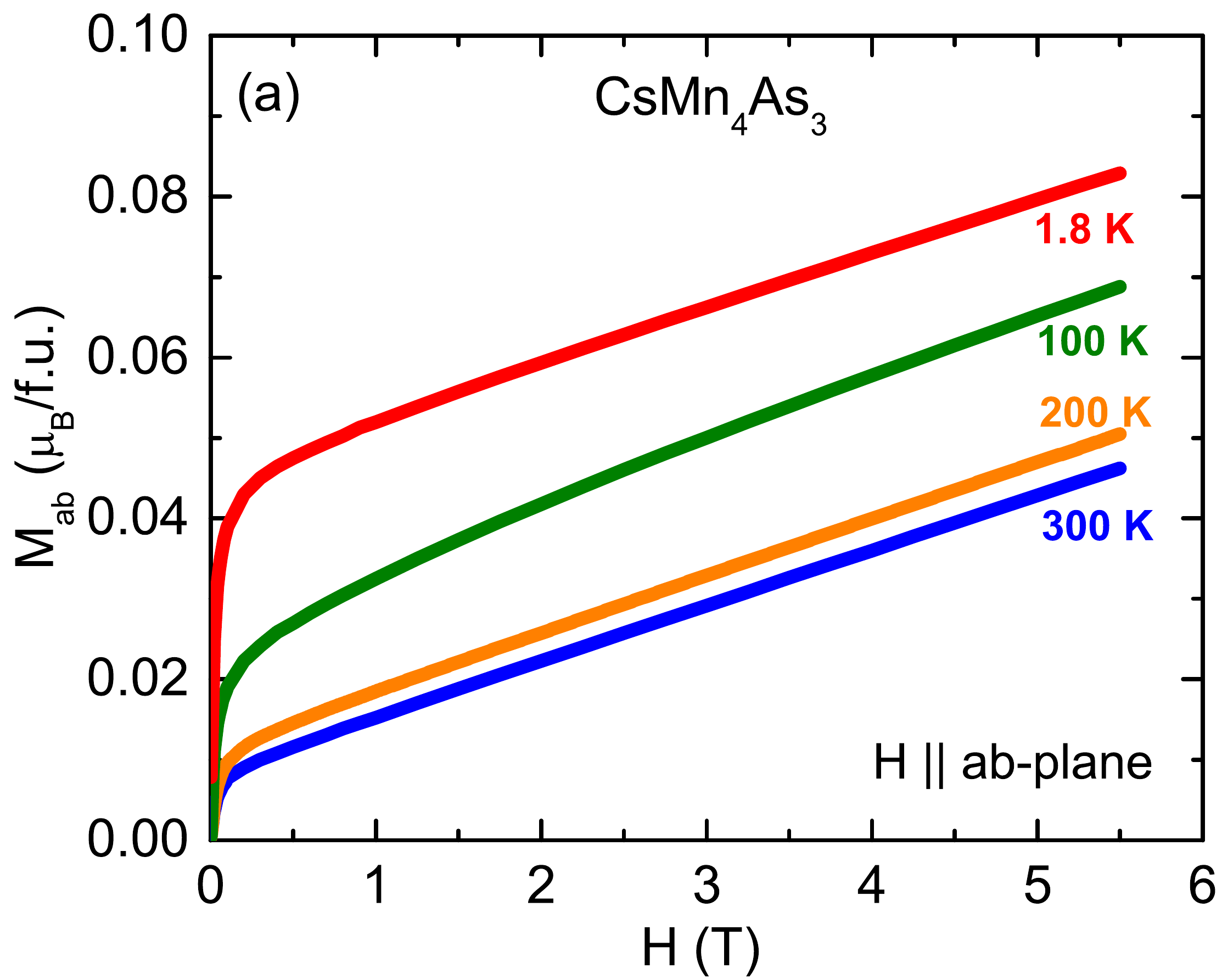}
	\includegraphics[width=3.3in]{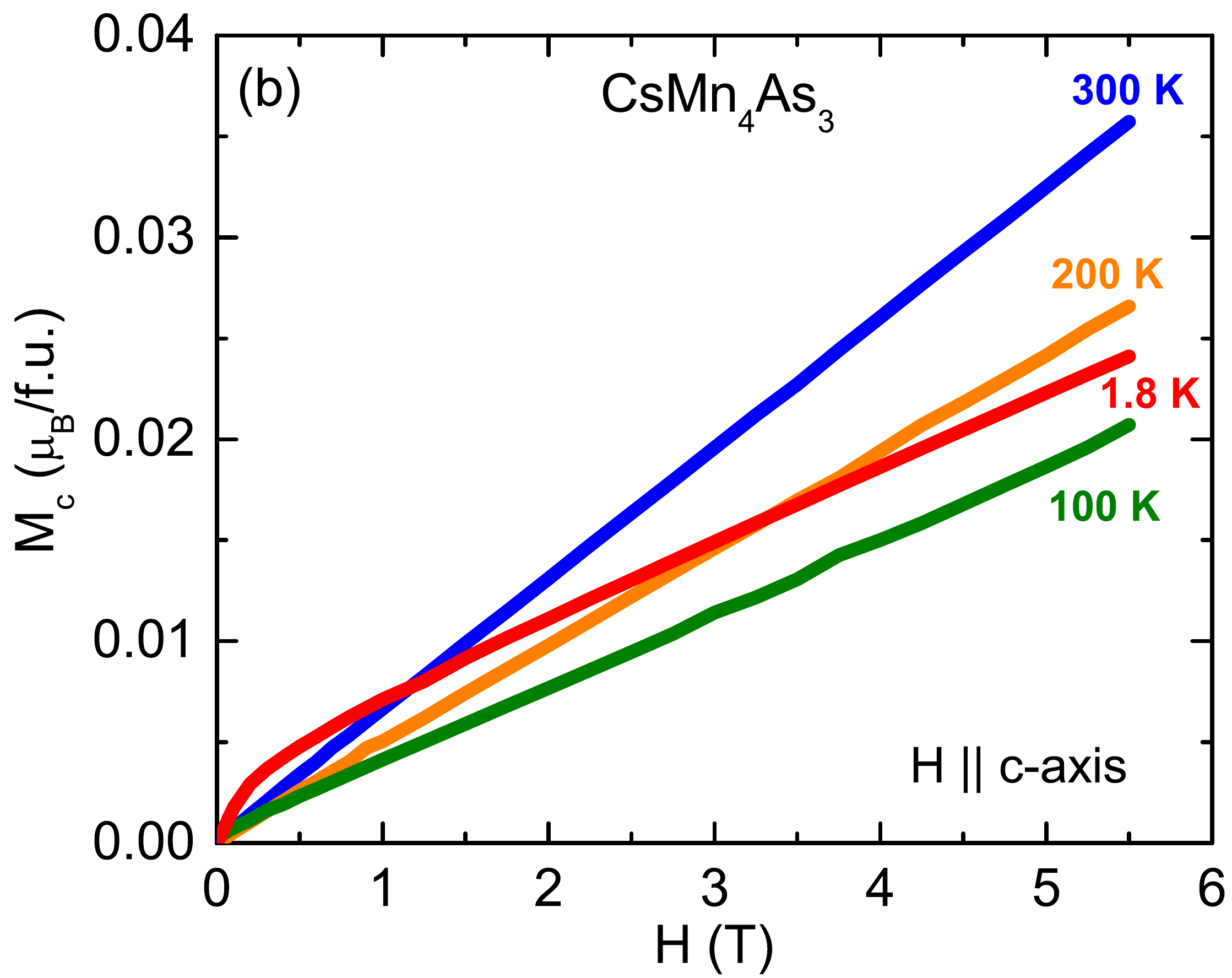}
	\caption{Isothermal magnetization $M$ versus applied magnetic field $H$ data of CsMn$_4$As$_3$ taken at four temperatures between 1.8 and 300~K for $H$ along the (a) $ab$-plane and (b) $c$-axis.}
	\label{fig:MH}
\end{figure*}
{\bf Magnetism.} The anisotropic magnetic susceptibilities $\chi_{ab,c} \equiv M_{ab,c}/H$ of CsMn$_4$As$_3$ measured in an applied dc magnetic field $H = 0.1$~T are shown in Fig.~\ref{fig:MT} (also see Fig.~S6, supporting information for corresponding data at $H = 3.0$~T). Some noticeable features observed in the $\chi(T)$ data are the following; (i) $\chi_{c}$ decrease monotonically with decreasing $T$ before showing an upturn at $T\sim 125$~K, (ii) $\chi_{ab}$ increases monotonically with decreasing $T$ and exhibits the features that suggest the presence of two magnetic transitions at $T_{1} = 325(5)$ and $T_{2} = 150(5)$~K, respectively. 

Since the CsMn$_4$As$_3$ crystals were grown using the MnAs flux which is a ferromagnet with $T_{\rm c} = 313(5)$~K\cite{Bean-1962, Haneda-1977, Maki-1998}, it's natural to consider that the transition observed at 325~K could possibly be an outcome of the inclusion of MnAs impurity in the crystals. However, the absence of this transition in the $\chi_{c}(T)$ data taken on the same crystal as well the jump observed in the $C_{\rm p}(T)$ at $T_{1}$ (Fig.~\ref{fig:HC}) together suggest that this feature is likely a bulk effect intrinsic to the material and is not due to ferromagnetic impurity. Additionally, NiAs-type hexagonal crystal structure and significantly different lattice parameters of MnAs \cite{Wilson-1964} compared to those of CsMn$_4$As$_3$ reduce the possibility of epitaxial growth of films of the former between the layers or on the surface of the latter, which is required for exhibiting the anisotropic impurity effect as observed in this case. Furthermore, the results obtained on the crystals of similar tetragonal materials grown using the same synthesis process suggest that the inclusion of MnAs impurity produces nearly isotropic effects in the magnetic measurements \cite{}(Fig.~S7, supporting information).

Although neutron diffraction measurements are needed to solve the underlying magnetic structure of this new material, we can still extract some features of the magnetism from the thermodynamic $\chi(T)$ and $M(H)$ measurements. Considering the fact that CsMn$_4$As$_3$ is a small band-gap semiconductor as well as from an analogy with the related BaMn$_2$As$_2$, it is likely that the magnetism of this material is of the local moment character where the Mn ion carries the magnetic moment. The $\chi_{c}(T)$ of CsMn$_4$As$_3$ that exhibits a behavior qualitatively similar to the BaMn$_2$As$_2$ \cite{Pandey-2013, Pandey-2015} suggests an AFM ground state in this material. The $\chi_{c}(T)$ data that approach to $\chi_{c} \approx 0$ for $T \rightarrow 0$ (Inset, Fig.~\ref{fig:MT}) when extrapolated from higher temperatures ($T>125$~K) suggests that the $c$-axis is the magnetic easy axis for the localized Mn-spins which are aligned antiparallel along this direction. The upturn observed in the $\chi_{c}(T)$ at 125~K is likely due to a minor misalignment of the crystal with respect to the $H$ \cite{Pandey-2013, Pandey-2015}. The two transitions observed in $\chi_{ab}$ at $T_1$ and $T_2$ likely arise from canting of the localized Mn-spins aligned along the tetragonal $c$-axis towards the $ab$-plane. The canting of the moment toward the $ab$-plane likely arises from a lattice distortion giving rise to an antisymmetric Dzyaloshinskii-Moriya interaction\cite{Glasbrenner-2014} and similar to the magnetic transitions in CaFe$_4$As$_3$,\cite{Zhao-2009, Nambu-2011} it could also involve a commensurate $\leftrightarrow$ incommensurate transition. 

As evident from the isothermal magnetization $M_{ab}$ versus $H$ measurements, these two successive canting events lead to small FM moments of $\mu_{1\,ab} \approx 0.01~\mu_{\rm B}$ and  $\mu_{2\,ab} \approx 0.045~\mu_{\rm B}$/Mn at 300 and 1.8~K, respectively [Fig.~\ref{fig:MH}(a)]. If we assume that the moment associated with the localized Mn-spins in CsMn$_4$As$_3$ are of the similar size ($\mu_{\rm Mn} \approx 4~\mu_{\rm B}$) of that in BaMn$_2$As$_2$ as indicated by the above band structure calculations, then the canting angles $\theta_1$ and $\theta_2$ associated with the canting events occurring at $T_1$ and $T_2$, respectively, can be calculated as 
\begin{equation}
\mu_{\rm Mn }{\rm cos}(90^{\circ}-\theta_{n}) = \mu_{n\,ab}/2.
\label{eq:Canting}
\end{equation}

Using eq~\ref{eq:Canting}, we estimate the canting angles as $\theta_{1} = 0.07^{\circ}$ and $\theta_{2} = 0.32^{\circ}$. 

Variations of $M_{c}$ with $H$ are shown for four temperatures in Fig.~\ref{fig:MH}(b). At high temperatures ($T>100$~K), following a behavior expected from an AFM material the $M_{c}$ varies linearly with $H$. However, for $T\le 100$~K, the $M_{c}(H)$ data start deviating from the linear $H$ dependence and develop a very small ordered moment $\mu_{c} \approx 0.004~\mu_{\rm B}$ at 1.8~K\@. As discussed before, the observed small ordered moment likely originates from a $\approx 4.8^{\circ}$ misalignment of the tetragonal $c$-axis with  $\bf{H}$.

\begin{figure*}
	\includegraphics[width=5.0in]{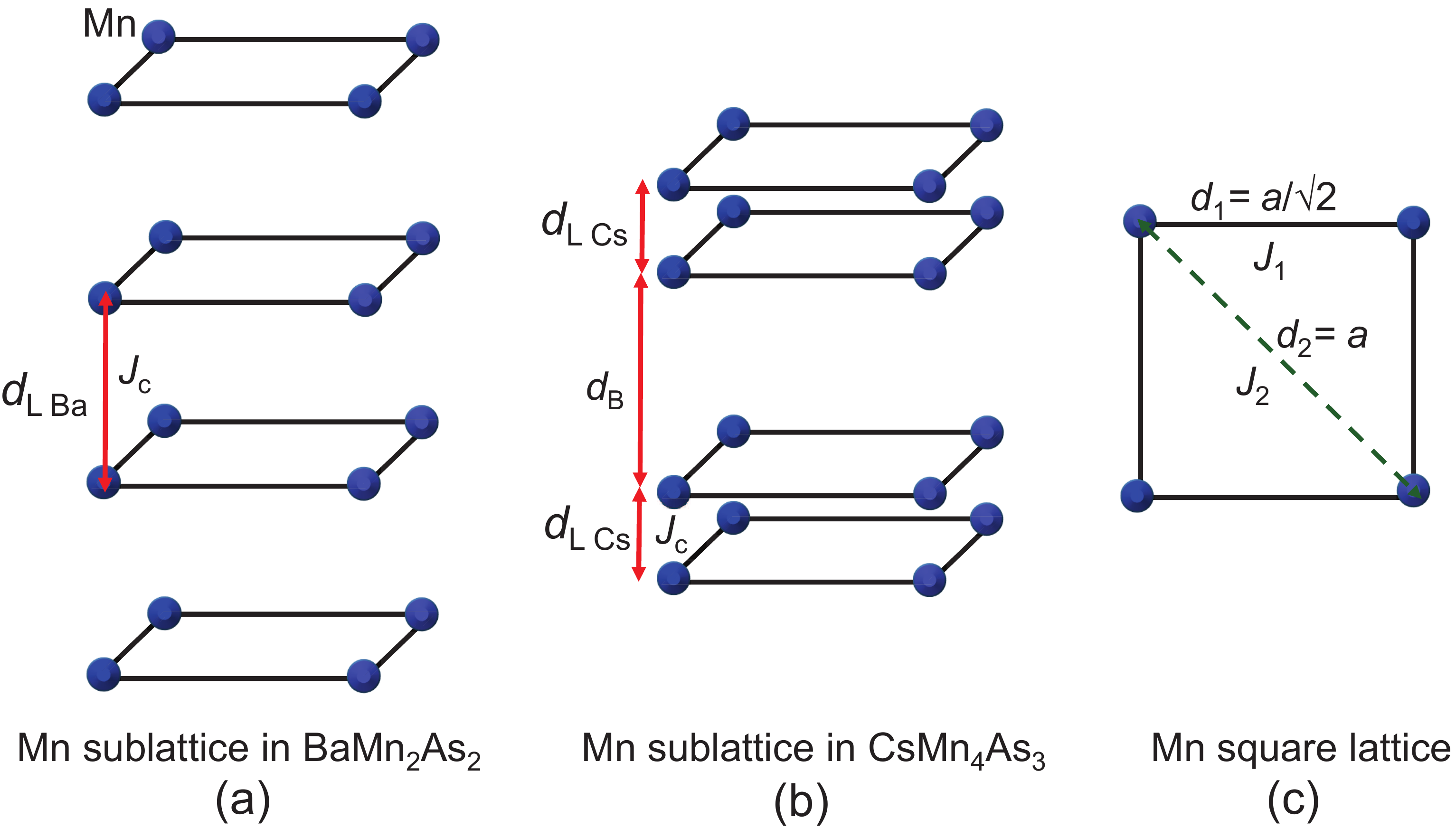}
	\caption{Representations of the Mn sublattices in (a) BaMn$_2$As$_2$ and (b) CsMn$_4$As$_3$. The inter-layer distances $d_{\rm L\,Ba}$, $d_{\rm L\,Cs}$ and $d_{\rm B}$ are shown on a proportional scale. (c) The top view of the Mn square lattice. Intralayer ($J_{1}$ and $J_{2}$) and interlayer ($J_{c}$) magnetic exchange interactions are indicated in each panel.}
	\label{fig:Mn-Lattice}
\end{figure*}

Although, apart from the canting of the moment toward the $ab$-plane, the magnetic characteristics of CsMn$_4$As$_3$ appear qualitatively similar to BaMn$_2$As$_2$ the Mn sublattice is very different in both the compounds. While in BaMn$_2$As$_2$ the layers of Mn square lattice are equidistantly placed along the crystallographic $c$-direction with the interlayer distance $d_{\rm L\,Ba} = c(1-2z_{\rm Mn}) = 6.7341(4)$~\AA, the Mn-atoms form bilayers in CsMn$_4$As$_3$ (Fig.~\ref{fig:Mn-Lattice}). Within a bilayer the interlayer distance is $d_{\rm L\,Cs} = c(1-2z_{\rm Mn}) = 3.1661(6)$~\AA, which is only about 40\% of the distance $d_{\rm B} = 2cz_{\rm Mn} = 7.290(6)$~\AA\ of the next-nearest layer that belongs to the adjacent bilayer. In fact, the $d_{\rm L\,Cs}$ in CsMn$_4$As$_3$ is nearly equal to the intralayer nearest neighbor (NN) distance $d_{1} = a/\sqrt{2} = 3.0322(2)$~\AA\ and 26\% smaller than the intralayer next nearest neighbor (NNN) distance $d_{2} = a = 4.2882(2)$~\AA\ (Fig.~\ref{fig:Mn-Lattice}). These significant differences between the Mn sublattices of CsMn$_4$As$_3$ and well-known Ba$M_2$As$_2$ ($M$: Cr, Mn, Fe, Co, Ni, Cu, Ru, Rh and Pd) compounds are bound to change the strengths of the exchange interactions as well as their interplay in the former compared to the latter within the $J_{1}\mhyphen J_{2} \mhyphen J_{c}$ model\cite{Johnston-2011}. Thus, this new material lays the foundation of a new playground where systematic investigations in the search of exciting electronic and/or magnetic properties can be performed by doping and pressure studies.

\section{Summary and Conclusions}
We have discovered a new layered tetragonal ternary compound CsMn$_4$As$_3$ (structure type: KCu$_4$S$_3$, space group: $P4/mmm$, No. 123 and $Z = 2$). Our physical property measurements show that similar to the related ThCr$_2$Si$_2$-type tetragonal BaMn$_2$As$_2$ this compound is a small band-gap semiconductor, thus the underlying magnetism very likely originates from localized Mn spins. Our magnetic measurements suggest that this compound has an AFM ground state with $T_{\rm N} > 300$~K and the magnetic easy axis is likely along the crystallographic $c$-axis. The magnetic susceptibility and magnetization data within the basal $ab$-plane show the signature of two magnetic moment canting events at $T_{1} = 325(5)$ and $T_{2} = 150(5)$~K with a low-$T$ canting angles $\theta_{1} \approx 0.1^{\circ}$ and $\theta_{2} \approx 0.3^{\circ}$, respectively. While the transition at $T_{2}$ almost certainly is intrinsic to the material, further work is needed to categorically resolve whether or not the transition at $T_{1}$ is intrinsic or is an artifact of epitaxially grown FM impurities between the layers or on the surface of CsMn$_4$As$_3$. Although the magnetic properties of CsMn$_4$As$_3$ appear qualitatively similar to the well-studied BaMn$_2$As$_2$, the underlying Mn sublattice is very different. While in BaMn$_2$As$_2$ the layers of the Mn square lattice are equidistant along the $c$-axis with interlayer distance $d_{\rm L\,Ba} = 6.7341(4)$~\AA, the Mn ions form a bilayer in CsMn$_4$As$_3$ with the interlayer distance within the bilayer $d_{\rm L\,Cs} =  3.1661(6)$~\AA\ and next-nearest interlayer distance between the adjacent bilayers $d_{\rm B} = 7.290(6)$~\AA. This modification leads into a situation where $d_{\rm L\,Cs}$ is approximately equal to the NN distance $d_{1}$ as well as smaller than the NNN distance $d_{2}$ within the Mn square lattice. These variations in the underlying Mn sublattice likely alter the energy balance between the exchange interactions within the $J_{1}\mhyphen J_{2} \mhyphen J_{c}$ model compared to those in BaMn$_2$As$_2$. 

This new 143-type addition to the family of layered tetragonal compounds related to Fe-based superconductors brings exciting opportunities to look for interesting magnetic as well as plausible superconducting ground states. While neutron diffraction and scattering experiments are needed to solve the magnetic structure as well as to determine the estimates of the strengths of the exchange interactions, doping and pressure studies with the aim of discovering new superconductors or other interesting ground states might be a fruitful endeavor. \\

{\bf Supporting Information.} Figures showing the powder x-ray diffraction data and their refinement, nearest-neighbor environments around Cs, Mn, As1 and As2 atoms in CsMn$_4$As$_3$, spin alignment in the A-type antiferromagnetic model used for the band structure calculations, estimation of magnetic entropy, evaluation of electrical resistivity data in the context of variable range hopping model, magnetic susceptibility versus temperature data of CsMn$_4$As$_3$ measured at $H = 3:0$~T, effect of MnAs impurity on the susceptibility of the related compound Ba$_{0.6}$K$_{0:4}$Mn$_2$As$_2$, table containing the anisotropic displacement parameters of Cs, Mn, As1 and As2 atoms estimated from single-crystal x-ray diffraction data of CsMn$_4$As$_3$, and a crystallographic information file.

\begin{acknowledgement}
The research at Ames Laboratory was supported by the U.S. Department of Energy, Office of Basic Energy Sciences, Division of Materials Sciences and Engineering. Ames Laboratory is operated for the U.S. Department of Energy by Iowa State University under Contract No. DE-AC02-07CH11358.
\end{acknowledgement}

\bibliography{CsMn4As3_Manuscript_IC}

\begin{tocentry}
	\includegraphics[width=3.3in]{Structure-1.pdf}\\
	
{\bf Synopsis:} Two layers of [Mn$_2$As$_2$]$^{2−}$ in BaMn$_2$As$_2$ are condensed into a single [Mn$_4$As$_3$]$^{−}$ layer in the new compound CsMn$_4$As$_3$ by removal of the intermediate layer of the cation as a consequence of the weak Cs-As bonding.  
\end{tocentry}
\end{document}